\documentclass[twocolumn]{revtex4}
\usepackage[dvips]{graphicx}
\usepackage{dcolumn}
\usepackage{amsmath}
\usepackage{bm}
\begin{document}
\title{Development of signal-extraction scheme 
for Resonant Sideband Extraction}

\author{K. Kokeyama}
\affiliation{The Graduate School of Humanities and Sciences, 
Ochanomizu University\\2-1-1, Otsuka, Bunkyo-ku, Tokyo, 112-8610 Japan}

\author{K. Somiya}
\affiliation{Max-Planck-Institut f\"ur 
Gravitationsphysik\\ Am M\"uhlenberg 1, 14476 Potsdam, Germany}

\author{F. Kawazoe}
\affiliation{The Graduate School of Humanities and Sciences, 
Ochanomizu University\\2-1-1, Otsuka, Bunkyo-ku, Tokyo, 112-8610 Japan}

\author{S. Sato}
\affiliation{TAMA project, National Astronomical 
Observatory of Japan\\2-21-1, Osawa, Mitaka, Tokyo 181-8588 Japan}

\author{S. Kawamura}
\affiliation{TAMA project, National Astronomical 
Observatory of Japan\\2-21-1, Osawa, Mitaka, Tokyo 181-8588 Japan}

\author{A. Sugamoto}
\affiliation{Ochanomizu University\\2-1-1, Otsuka, Bunkyo-ku, Tokyo, 112-8610 Japan}
\begin{abstract}
 As a future plan, an advanced gravitational-wave detector
will employ an optical configuration of resonant sideband extraction (RSE),
achieved with an additional mirror
at the signal-detection port of the power-recycled
Fabry-Perot Michelson interferometer.
To control the complex coupled cavity system,
one of the most important design issues is how to extract
the longitudinal control signals of the cavities.
We developed a new signal-extraction scheme
which provides an appropriate sensing matrix.
The new method uses two sets of sidebands:
one of the sideband components satisfies the critical coupling condition
for the RSE interferometer and reaches the signal-extraction port,
and the other sideband is completely reflected
by the Michelson interferometer.
They provide a diagonalized sensing matrix
and enable the RSE control to be robust.
\end{abstract}
\maketitle
\section{introduction\label{1}}

The gravitational-wave detection will provide the evidence
of the most important part of Einstein's general theory of relativity
and will start a new era of observational astronomy. 

Gravitational waves are the ripples of space-time
that travel through space.
They have extremely little interaction with matter,
but they cause the differential displacement of free masses.
Around the world, several interferometric gravitational-wave observatories aim
at the first detection of gravitational waves.
The ground-based gravitational-wave antennas are
based on a Michelson interferometer (MI) with a stabilized laser.
When the gravitational waves pass, they will produce
a differential length change of the interferometer's arms.
The displacement will be detected as a relative phase change
of these two arms.  However, the magnitude of the displacement
measured by the detectors is so small, only about $10^{-20}$m,
even for large scale detectors,
that the sensitivity is disturbed by various kinds of noise.
Toward the achievement of extremely high sensitivity,
several optical configurations and various noise reduction techniques
have been developed. But so far, no attempt to directly detect
gravitational waves has been successful.

The first-generation detectors such as TAMA300 ~\cite{TAMA},
VIRGO ~\cite{VIRGO}, GEO 600 ~\cite{GEO},
and LIGO~\cite{LIGO1, LIGO2} have been constructed and are currently in operation.
The power-recycled Fabry-Perot (FP) Michelson interferometer
is employed as the optical configuration for most of these detectors.
Two FP resonant cavities are placed to increase the light storage time
and enhance the gravitational wave signals.
In a simple MI case, the interferometer is kept on a dark fringe
at the detection port@to optimize the shot-noise-limited sensitivity
and all the carrier light will return to the laser and will be wasted.
To utilize these lights effectively,
a power-recycling mirror (PRM) is placed in front of the MI
so that the effective laser power is increased in the interferometer.
\begin{figure}
\begin{center}
\includegraphics[width=8cm]{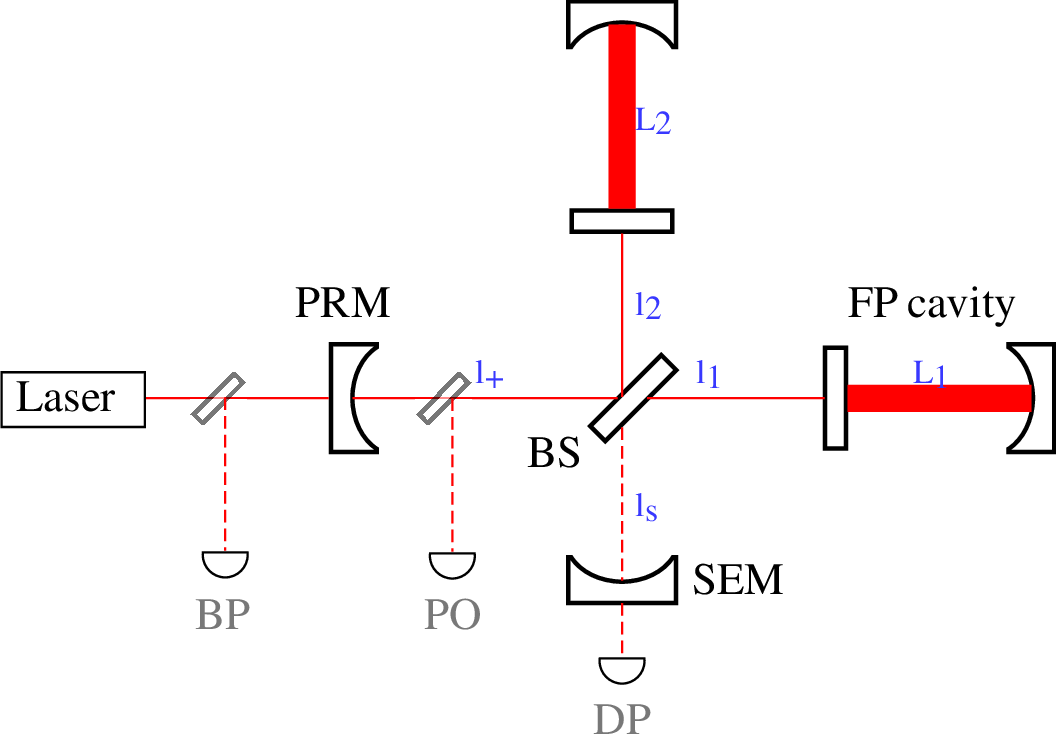}
\vspace{5mm}
\caption{Optical configuration of RSE. Two FP cavity lengths are
$L_1$ and $L_2$. The MI has two light paths
of length $l_1$ and $l_2$. $l_{\rm p}$ is the distance
from the PRM to the BS and $l_{\rm s}$ is that from the BS
to the SEM.
There are three detection ports for length-sensing:
the BP, the DP and the PO.}
\label{fig:RSE configuration}
\end{center}
\end{figure}
%
Although these detectors are in the process of reaching
a remarkable sensitivity,
current-operating detectors can detect
the gravitational wave events only in a range of about 15 Mpc at the maximum.
This is not enough to establish gravitational wave astronomy.

More sensitive second generation detectors are being planned and developed.

The Japanese future plan,
Large Cryogenic Gravitational-wave Telescope (LCGT)~\cite{LCGT}
and the US plan, Advanced LIGO \cite{adLIGO1} are undergoing development
as the second generation gravitational-wave detectors.
Their sensitivities will improve
with various new techniques such as very high power laser,
advanced suspension systems, cryogenics, new optical configurations, etc.
As one of these advanced techniques,
the resonant sideband extraction (RSE) \cite{RSE} is selected
as the standard optical configuration of these future detectors.
The RSE topology requires an additional mirror at the dark port.
However, this mirror adds another longitudinal degree of freedom
to control and makes its optical configuration complex.
Therefore, it is essential to have a length-sensing scheme suitable
for the RSE interferometer.

In the next section, we will review the RSE interferometer,
in particular, focusing on its optical configuration.
Then we will propose a new length-sensing scheme in Section \ref{3},
and explain how it developed in Section \ref{4}.
In Section \ref{5}, we will show the simulation result
and the discussion. At the end, we will summarize our work.

\section{length sensing of RSE\label{2}}

The optical configuration of the RSE interferometer is shown
in Figure \ref{fig:RSE configuration}.
The interferometric part consists of
two input test masses and a beam splitter (BS).
An input test mass and an end test mass form an FP cavity
which plays an important role
in the gravitational-wave signal enhancement.
Additionally, the interferometer also has a PRM at the bright port (BP)
\cite{footnote1} and a signal extraction mirror (SEM) at the dark port (DP).

Since the arm cavities have a high finesse
(which indicates how many round trips light travels in a cavity)
in the RSE configuration, the gravitational-wave signals are enhanced accordingly.
However, the signals are over-circulated and they are canceled in the arm cavity
owing to the phase changes of the gravitational wave signals.
The SEM plays an important role in circumventing this problem.
The signal extraction cavity (SEC) lowers the the effective finesse
for the gravitational-wave signals and they escape from the cavities
before over-circulation. On the other hand, the entire carrier light returns
to the BP, so the SEC does not affect the power in the arm cavities.
Thus the sensitivity will be improved.

The advantage of the RSE configuration is that
the thermal problems can be circumvented  being compared
with an optical configuration of FP
MI with a power-recycling cavity (PRC), which has
the same sensitivity.
The light power at the beam splitter or at the PRC is lower
in the RSE case than in the FP MI case with a PRC.
Especially for the LCGT, the mirrors will be kept at a super cryogenic temperature,
thus it is necessary that the heat produced by the laser light absorption
in the bulk of the input test masses is released through the suspension systems.
If the heat is too high, the suspension systems cannot be refrigerated.
Furthermore, even if the beam splitter and the PRM are not
cooled, the thermal lens effect could cause.
It would disrupt the mode matching of the light field
and decrease the power in the arm cavities,
possible leading to a worse shot noise limit.

To operate the interferometer as the gravitational-wave detector,
all the mirrors have to be controlled at the proper positions
so that the cavities are on resonance for the light fields.
The longitudinal signals should be sent to the mirrors
to control their positions. For the RSE interferometer,
as shown in Fig.~\ref{fig:RSE configuration},
there are five degrees of freedom to control;
the arm-length common mode $L_+$, the differential mode $L_-$,
the PRC length $l_+$,
the MI differential length $l_-$, and the SEC length $l_{\rm s}$.
The exact definitions of these
lengths are shown in table~\ref{5length}.

$L_+$ and $L_-$ signals are extracted relatively independently
since the phase sensitivities are enhanced by its stored light fields
in the high finesse arm cavities. The difficulty is to extract the three signals
of the central part of the RSE. If there is no cross talk between these signals,
only one signal component can be sent and fed back to one degree
of freedom most appropriately. In this ideal control condition,
the system will be very robust. However, 
in most of the control schemes currently used or planned for use,
these three signals are mixed with the others.

\begin{table}
\begin{center}
\begin{tabular}{l|c|c}
    \multicolumn{3}{c}{}	\\ \hline
Description               & Symbol & Length \\ \hline \hline
Common arm cavity    & $L_+ $& $L_1 + L_2$ \\
Differential arm cavity & $L_-$ & $L_1 - L_2$ \\
PRC                 & $l_+ $ & $l_{\rm p} + \bar{l} $ \\
Differential MI 			& $l_- $ & $l_1 - l_2$ \\
SEC                  & $l_{\rm s} $ & $l_{\rm s} + \bar{l} $ \\ \hline   
 \end{tabular} 
 \end{center}
\label{5length}
\caption{Five longitudinal degrees of freedom of the RSE.
$L_1$ and $L_2$ denote the inline and perpendicular arm lengths.
$l_1$ and $l_2$ are the inline and perpendicular path lengths of a MI.
$\bar{l} = (l_1+l_2) /2$; the average light path length of MI.
$l_+$ and $l_{\rm s}$ designate the distance from the PRM to the BS,
and the SEM to the BS.
They are also found in Fig.~\ref{fig:RSE configuration}.}
\end{table}

\section{\label{merged method}Coupled - reflected method\label{3}}

\subsection{Definition of coupled - reflected method}
Our scheme, {\it critically coupled - reflected method}
({\it coupled - reflected method} for short)
employs two sets of rf sidebands which do not enter
the arm cavities to prevent the admixture of 
the arm length signals ($L_+$ and $L_-$) and
the other signals ($l_+, l_-$ and $l_{\rm s}$).
In the conventional way, $L_+$ and $L_-$ signals are extracted 
by beating between the carrier light and PM sidebands
at the BP and DP, respectively.
%
To obtain $l_+, l_-$ and $l_{\rm s}$ signals,
the two sets of sidebands are double-demodulated at the PO.
This technique requires that one of the two sidebands is phase-modulated 
whereas the other sideband is amplitude-modulated to be a local field
against the PM field on the double demodulation scheme.
The resonant conditions of the two sets of sidebands are;
\begin{description}
\item{(i)} The PM sidebands reach the SEM by the critical coupling condition
to carry information of the SEC effectively~\cite{low}.
The PM fields are resonant inside the PRC and the SEC.
\item{(ii)} The AM sidebands are reflected completely by the MI
so as not to carry any information of the SEC length.
This condition can prevent a complex mixture between $l_{\rm s}$ and $l_+$.
\item{(iii)} The AM sideband is an applied so-called delocation scheme.
Delocation is macroscopic detuning of AM sidebands by a slight change
in the PRM position~\cite{sato1}.
The length-sensing matrix can be diagonalized by 
this delocation scheme.
\end{description}

\begin{figure}
\begin{center}
\includegraphics[width=7cm]{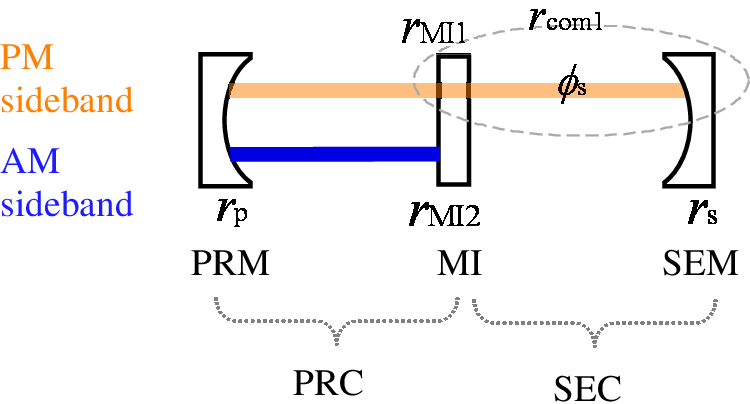}
\vspace{3mm}
\caption{An RSE interferometer can be modeled by a coupled cavity
which contains the PRM (reflectivity $r_{\rm p}$),
the MI and the SEM (reflectivity $r_{\rm s}$).
The former cavity corresponds to the PRC,
and the second one corresponds to the SEC.
$r_{\rm MI1}$ and $r_{\rm MI2}$ are the reflectivities
of the MI for the PM and AM sidebands respectively.
$r_{{\rm com}1}$ is the reflectivity of the compound mirror
composed of the MI and the SEM, for the PM sideband.
}
\label{fig:coupled cavity}
\end{center}
\end{figure}

The pair of sideband frequencies have to be determined
to satisfy conditions (i) and (ii).
To make the issue less complicated,
let us introduce a view of the RSE interferometer
as a coupled cavity system.
The central part of the RSE interferometer can be modeled by a coupled cavity
which contains three mirrors (see figure \ref{fig:coupled cavity}).

%

The reflectivity of the middle compound mirror
(which corresponds to the MI), $r_{\rm MI}$, can be determined
by choosing the MI asymmetry length, $\Delta \ell$,
and the sideband frequency, $f_j$;
\begin{align}
&r_{{\rm MI}j} = \cos \alpha _j\\
\label{eq:asy1}
&\alpha _j= \frac{\Delta \ell~2 \pi f_j} {c}
\end{align}
where $j=1, 2$ for the PM and AM sidebands respectively.
The reflectivities of the arm cavities are assumed
since neither of the sidebands go into the cavities.

We shall consider that the middle mirror and the SEM make
another compound mirror again. 
The reflectivity of the new compound mirror is
\begin{equation}
\label{eq:asy4}
r_{{\rm com}j} = \cos \alpha _j 
-\frac{r_s \sin ^2 \alpha _j e^{i \phi_{{\rm s}j}}}
{1-r_s \cos \alpha _j e^{i \phi_{{\rm s}_j}}},
\end{equation}
where $\phi _{{\rm s}j}$ is the round trip phase
between the middle mirror and the SEM.

\begin{figure}
\begin{center}
\includegraphics[width=8cm]{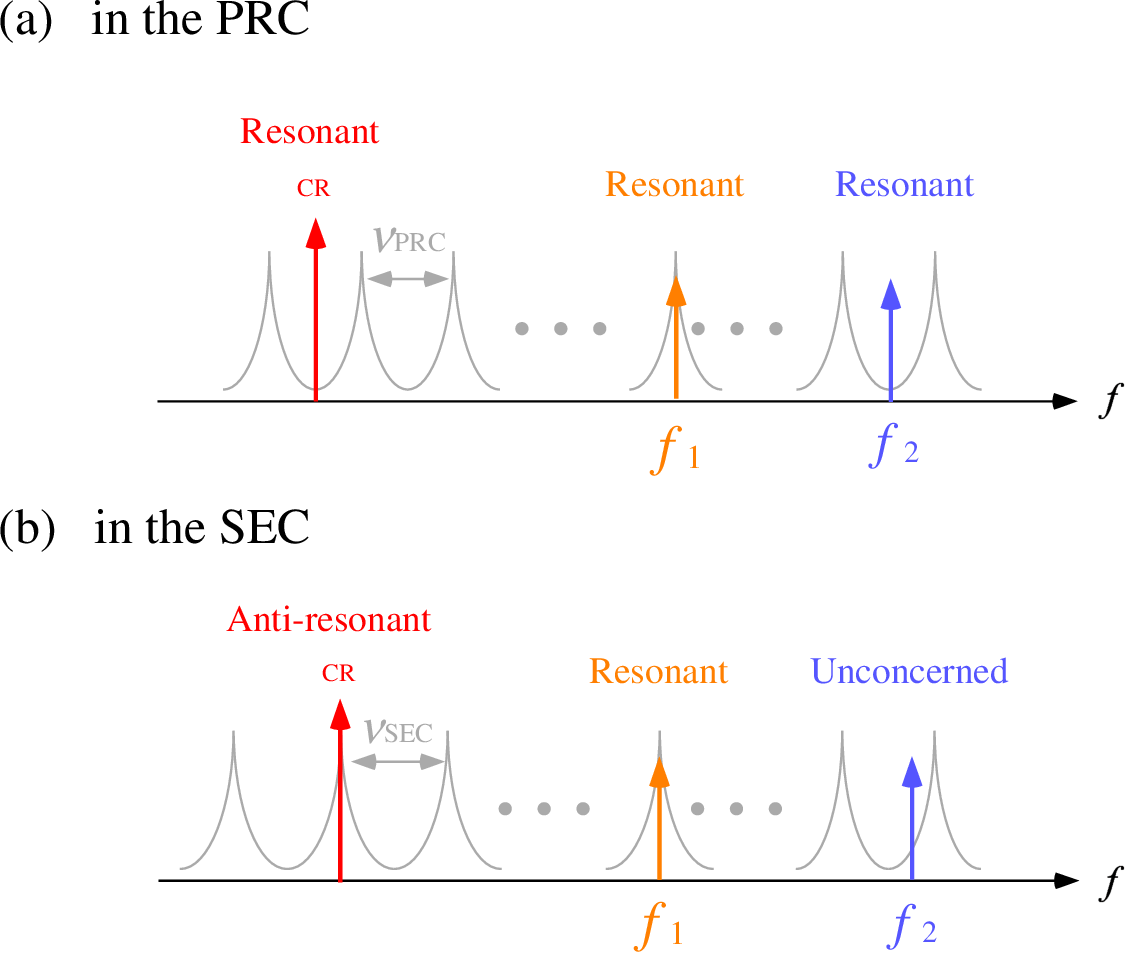}
\vspace{2mm}
\caption{
Relation between the cavity FSRs and the sideband frequencies.
(a) PRC transmission curve.
The FSR of the PRC is $\nu_{\rm PRC}$.
The carrier light (CR) is resonant
since the sign is flipped by the arm cavities.
The PM sidebands ($f_1$) are resonant.
The AM sidebands ($f_2$) are resonant
since the sign is flipped by $r_{\rm MI2}=-1$.
(b) SEC transmission curve.
The FSR of the SEC is $\nu_{\rm SEC}$.
The carrier light is anti-resonant.
The PM sidebands are resonant.
The AM sidebands do not depend on $\nu_{\rm SEC}$
since it does not enter the cavity.
}
\label{fig:cavity reso}
\end{center}
\end{figure}

\begin{description}
\item{(i)} To determine the PM sideband frequency, $f_1$

PM sidebands satisfy the critical coupling condition.
In general, critical coupling is achieved
when the input and end mirror have the same reflectivity
and the laser lights are resonant in the cavity.
The laser fields pass through the cavity
without returning to the input port.
To apply this condition for $f_1$ to the RSE interferometer,
the PRM and the compound mirror which consists of the MI and SEM
have the same reflectivity.
Therefore, the reflectivity of the compound mirror satisfies,
\begin{equation}
\label{eq:rcom}
r_{{\rm com}1} = r_{\rm p}.
\end{equation}
$f_1$ should be resonant in the SEC, $e^{i\phi _{{\rm s}1}}=1$.
Substituting these conditions into Eq.~(\ref{eq:asy4}), we get

\begin{equation}
f_1=\frac{c}{2 \pi \Delta \ell} ~\cos ^{-1} \Bigl( \frac{r_{\rm p}+r_{\rm s}}
{r_{\rm p} r_{\rm s}+1} \Bigr)
\end{equation}

%
%
To realize the resonant condition in the PRC,
$f_1$ should also satisfy
\begin{equation}
f_1= \Bigl(N+\frac{1}{2} \Bigr) \nu_{\rm PRC} 
\label{eq:FSR1}
\end{equation}
where 
$\nu_{\rm PRC}$ is the free spectral range (FSR)
of the PRC, which is $\nu_{\rm PRC}=c/2\ell _{\rm PRC}$
and $N$ is an arbitrary integer.
This is because the sign of the carrier field is inverted
by coming back from the two arm cavities,
and they are resonant in the PRC
when the PRC length is the anti-resonant condition for the carrier light.
See, Fig.~\ref{fig:cavity reso} (a).
In a similar way, $f_1$ also satisfies
\begin{equation}
f_1=N' \times \nu_{\rm SEC} 
\label{eq:FSR2}
\end{equation}
where 
$\nu_{\rm SEC}$ is the FSR
of the SEC and $N'$ is an arbitrary integer.
As is shown in Fig.~\ref{fig:cavity reso} (b),
the SEC cavity length is set as the resonant for the carrier light.
With this condition, the carrier light is virtually anti-resonant
in the SEC because the carrier light comes back from the arm cavities.

\item{(ii)} To determine the AM sideband frequency, $f_2$

To make the AM sidebands be reflected by the MI,
$r_{\rm MI2} =\pm1.$ Here, we supposed
$r_{\rm MI2} =-1$ to make $f_2$ as low as possible.
Therefore the AM sideband frequency satisfies 
\begin{equation}
f_2=\frac{c}{2\Delta \ell}.
\label{eq:amfreq}
\end{equation}
In addition, $f_2$ also satisfies 
\begin{equation}
\label{eq:9}
f_2=\nu_{\rm PRC} \times N''
\end{equation}
to be resonant in the PRC.
$N''$ is an arbitrary integer.
This is because the sign of the AM sideband fields are inverted 
by $r_{\rm MI2}=-1$.
%
\end{description}
%
Practically, when we design a interferometer,
a so-called mode cleaner 
between the modulators and the interferometer
will be necessary to eliminate the higher order
Gaussian modes of the laser light.
In order both sets of sidebands and the carrier light
to transmit through the mode cleaner,
the values of $f_1$ and $f_2$ should be integer multiples
of the FSR of the mode cleaner cavity.
However, under this additional condition,
$f_1$ and $f_2$ cannot satisfy Eq.~(\ref{eq:FSR1}) and Eq.~(\ref{eq:9}) at the same time.
%
Therefore, in practical ways,
we may use the values which satisfy the above conditions approximately
and do not disturb the optical conditions.

This is one challenge of this method.
The suitable optical parameters are discrete
and not flexible for the changes of interferometer design
because the parameters such as mirror reflectivities, 
the asymmetry length of MI and sideband frequencies
depend on the optical design.

%

\subsection{Features of Coupled-reflected method}

One of the advantage of the coupled-reflected method is that
short asymmetry length can be achieved.
We can choose any asymmetry length as long as the frequency of AM sidebands satisfies Eq.~(\ref{eq:amfreq}).
There are two advantages of short asymmetry length:
easier mode matching and fewer phase noises.
The effort of mode matching is easier when the inline
and perpendicular path are identical.
And also fewer laser phase noises are expected.
The phase noises can be canceled at the DP
when the inline and the perpendicular
path has approximately the same length.

However, as a difficulty, the frequency $f_2$
tends to be high due to a very short asymmetry length,
though the AM sideband is not the local oscillator for the gravitational waves.
For example, these parameters assume that
the asymmetry length is 0.824m and $f_2=182$MHz
in our simulation for LCGT (see Table \ref{matrix}).


The second benefit of the coupled-reflected method is that 
the clean $l_-$ signals can be acquired at the DP.
In general, $L _-$ and $l_-$ should be isolated from
$L_+, l_+$ and $l_{\rm s}$ signals as much as possible,
because $L _-$ and $l_-$ may include the gravitational wave signals
and the two signals cannot be separately extracted in principle.
The coupled-reflected method allows the $l_-$ signals
without admixture of $l_+$ and $l_{\rm s}$
because there are no AM sidebands in the SEC
owing to the condition of $r_{\rm MI2}=-1$.

Another major advantage of the coupled-reflected method
is that the delocation technique can
diagonalize the sensing matrix optically.
When one of the $l_+, l_-$, and $l_{\rm s}$ signals is extracted, 
two kinds of demodulation phases can be chosen
so that one desired signal is maximized,
or an undesired signal is minimized.
The $l_+, l_-$, and $l_{\rm s}$ signals are superposing
in the demodulation-phase domain
and cannot be extracted individually.
Especially the $l_+$ and $l_{\rm s}$ signals have the exact
same dependencies on the demodulation phases.
The delocation scheme can avoid this degeneracy.
The delocation is a macroscopic detuning for the AM sidebands
by changing the position of the PRM.
The off-resonant AM fields can change 
the optimum double-demodulation phases for the three signals.
The appropriate delocation amount realizes
that one desired signal can be extracted
while the undesired signals are zero
on a pair of appropriate delocation phases.
Therefore, the exact diagonalization
of the sensing matrix is possible 
when optimum demodulation phases are chosen.
Although the PRM position change affects
the resonant condition of the PM sideband field
in the PRC as well as the AM sidebands,
the PM sidebands are only slightly off resonant
since the delocation phase for the PM sidebands are small.
This is because the delocation phase is proportional to the sideband
frequencies and $f_1$ is much smaller than $f_2$.
See Eq.~(\ref{delopha}) in Section \ref{5}.
%
%
%
%

The delocation technique is seem to decrease about 40\%
of signal compared with no delocation.
This is because they are displaced from its resonant point
and the power of AM sideband fields is reduced in the PRC.

\section{Relation to other schemes\label{4}}

\begin{figure*}
\begin{center}
\includegraphics[width=13cm]{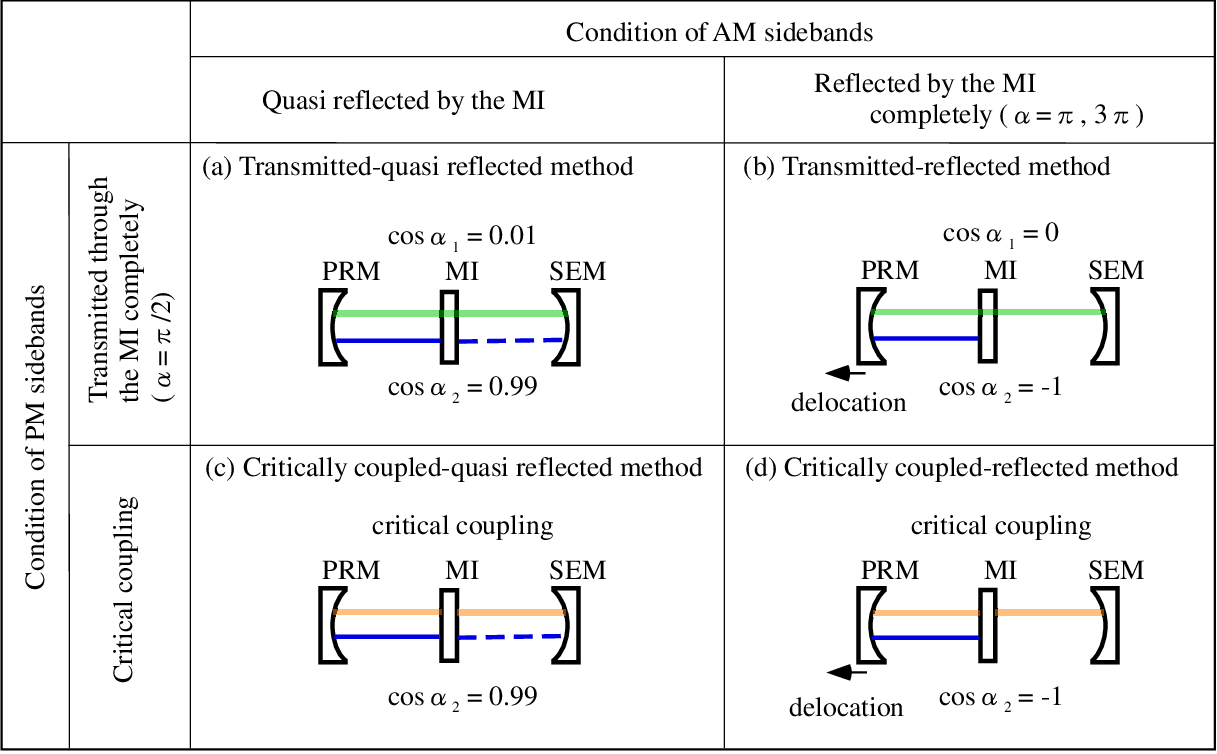}
\vspace{5mm}
\caption{Relation of four signal sensing and control schemes.
They are categorized by the optical conditions of the sidebands
in the PRC and SEC.
The PM sidebands are critically coupled or they transmit through the MI.
The AM sidebands are quasi-reflected or completely reflected by the MI.
The combinations of these conditions provide the four schemes.
}
\label{fig:history-table}
\end{center}
\end{figure*}

The coupled-reflected method was derived from three control schemes
for the RSE interferometer which has already been discussed,
or has been tested by prototype interferometers.
The schemes can be categorized 
from the viewpoint of the sideband options.
Depicted in Fig.~\ref{fig:history-table},
each set of sidebands has two options:
for PM sidebands, the critically coupled
or the complete transmission trough the MI;
for AM sidebands, the complete reflection
or the quasi-reflection by the MI.
The combination of these options derives four schemes,
the transmitted-quasi reflected method,
the critically coupled-quasi reflected method,
the transmitted-reflected method,
and the critically coupled-reflected method.
The critically coupled-reflected method was inspired as the fourth panel
of the table shown in Fig.~\ref{fig:history-table} (d).

The transmitted-quasi reflected method
was developed by the LIGO group as a default method
for Advanced LIGO \cite{adLIGO2}.
As is shown in Fig.~\ref{fig:history-table} (a),
almost all of the PM sidebands transmit the MI.
The frequency of the PM sidebands which satisfies
$r_{\rm com}=-r_{\rm p}$ is adopted for this method.
On the other hand, 
a great part of the AM sidebands is reflected by the MI.
The PM sidebands resonate both in the PRC and in the SEC,
and the AM sidebands resonate only in the PRC.
It is noted that the original scheme was not exactly
the same as the transmitted-quasi reflected method
since it was for a detuned RSE interferometer \cite{footnote2}.

The transmitted-reflected method has been developed by Sato
as the control scheme for LCGT \cite{sato1}.
As is shown in Fig.~\ref{fig:history-table} (b),
the PM sidebands pass through the MI completely
and all the AM sidebands are reflected by the MI.
The MI asymmetry length and the sideband frequencies
satisfy $\cos \alpha _1 = 0$, $\cos \alpha _2 = -1$
\cite{footnote} so that the sideband conditions are met.
The delocation technique is first introduced with this method.

The critically coupled-quasi reflected method
has already been tested by Somiya \cite{low}.
As is shown in Fig.~\ref{fig:history-table} (c),
the PM sidebands are critically coupled
and the AM sidebands slightly reach the SEM.
The critical coupling condition was first introduced
with a method to maximize $l_{\rm s}$ signals by this method.

Comparing these four methods,
one can find the coupled - reflected method 
inherits the advantages of other methods:
A short asymmetry length of MI can be available
inheriting the advantage of the critically
coupled - quasi reflected method;
no cross-talk between $L_+, l_+$ and $l_{\rm s}$
in the $l_-$ signals at the DP
inherits the advantage of the transmitted - reflected method;
the diagonalized length sensing matrix
by the delocation technique
inherits the advantage of the transmitted - reflected method as well.
In the next section, the analytical expressions
and the numerical simulation results
of the coupled - reflected method will be shown.

\section{Control signals and simulation result\label{5}}

The transfer function of light fields from the input 
to an arbitrary port is expressed as
\begin{equation}
T = \frac{E}{E_{\rm{in}}}
\end{equation}
where $E_{\rm in}$ is the light field of the incident beam
and $E$ is the light field  at the port.
$E_{\rm in}$ contains the carrier component
and two sets of upper ($\rm U$) and lower ($\rm L$) sideband components
which are generated by the EOMs.
The transfer function at the PO is
\begin{equation}
\label{eq:field}
T_{j} = \frac{t_{\rm p}}{1-r_{\rm p} R_{j} e^{-i\phi _{{\rm{p}}j}}}
\end{equation}
where $j=1, 2$ for the PM and AM sidebands, respectively
and $\phi_{{\rm p}j}$ is the round trip phase in the PRC.
$R_{j}~(j=1,2)$ is the reflectivity of the compound cavity
which includes the MI, FP arms and SEC
for the PM and AM sidebands, respectively;
\begin{align}
\nonumber
R_{j} = &r_{{\rm cav}j} \Biggl [ e^{-i \phi _{\rm{avr}}} \cos \alpha _{j} \\
&- \frac{e^{-2i \phi _{\rm {avr}}} \sin ^2 \alpha _{j} r_{\rm s} r_{{\rm cav}j}
e^{-i \phi _{{\rm s}j}}} {1- r_{\rm s} r_{{\rm cav}j}
\cos \alpha _{j} e^{-i \phi _{\rm{avr}}} e^{-i \phi _{\phi _{{\rm s}j}}}} \Biggr ]
\end{align}
where $\phi _{\rm{avr}}$ is the average round trip phase of $l_1$ and $l _2$
and $r_{{\rm cav}j}$ is the compound reflectivity of the arm cavity
for each sideband component.

%
The sensitivities of the double-demodulated signals at the PO,
$V_{\rm PO}$ for $l_k~(k=+,-, \rm{p})$ are given as 
\begin{align}
\label{eq:V}
\nonumber
\frac{\partial V_{\rm PO}}{\partial l_k}
=& \frac{\partial}{\partial l _k} \Re \bigl[ \{ T_{\rm U1}T^*_{\rm L2} + T_{\rm L1}^*T_{\rm U2} \} e^{-i(\delta 1+\delta 2)} \\
 & {}+ \{ T^*_{\rm U1} T_{\rm U2}
+ T_{\rm L1} T^*_{\rm L2} \} e^{-i(\delta 1-\delta 2)} \bigr]
\end{align}
where $T_{\rm{U1 (L1)}}$ represents the transfer functions of upper (lower) PM sidebands and $T_{\rm{U2 (L2)}}$ is the transfer functions of AM sidebands
at the PO. $\delta 1$ and $\delta 2$ are the demodulation phases.

From Eq.~(\ref{eq:field}) and (\ref{eq:V}),
the sensitivities of $V_{\rm{PO}}$ for $l_+,l_-$ and $l_{\rm s}$ are 
\begin{widetext}
\begin{align}
\label{eq:calc-p}
\frac{\partial V_{\rm{PO}}}{\partial l_{+}}
& \propto   \Im [i g_1 e^{-i \delta _1}] \Re[-g_2^2 R_2 e^{-i \triangle  _{\rm{p}2}}
e^{-i \delta _2}] + \Re [g_1^2 R_1 e^{-i \triangle  _{\rm{p}1}}
e^{-i \delta _1}]\Re[g_2 e^{-i \delta _2}]\\
\label{eq:calc-m}
\frac{\partial V_{\rm PO}}{\partial l _-} 
& \propto \Re[i g_1^2 e^{-i \triangle  _{\rm{p}1}} 
e^{-i \delta _1}] \Re[g_2 e^{-i \delta _2}]\\
\label{eq:calc-s}
\frac{\partial V_{\rm PO}}{\partial l_{{\rm s}}} 
& \propto -\Re[g_1^2 e^{-i \triangle _{\rm{p}1}} e^{-i \delta _1}] \Re[g_2 e^{-i \delta _2}]
\end{align}
\end{widetext}
%
%
where $g_{j}$ is a square root of the power recycling gain for each sideband;
\begin{eqnarray}
g_j= \frac{t_{\rm p}}{1-r_{\rm p} R_j e^{-i \triangle _{{\rm p}j}}}
\end{eqnarray}

\begin{eqnarray}
\label{delopha}
\triangle _{{\rm p}j} = \frac{2 \pi f_{j}l_{\rm del}}{c} .
\end{eqnarray}
These equations include the effect of the delocation.
The sidebands undergo delocation phase shifts, $\triangle _{{\rm{p}}j}$,
caused by the macroscopic length displacement, $l_{\rm del}$,
from the resonant point of the PRC.  The two sideband fields experience
different phase shifts because their phase changes depend on 
both the sideband frequencies and the displacement length.
The SEC does not experience any change by the delocation.

\begin{figure}[t]
\begin{center}
\includegraphics[width=10cm]{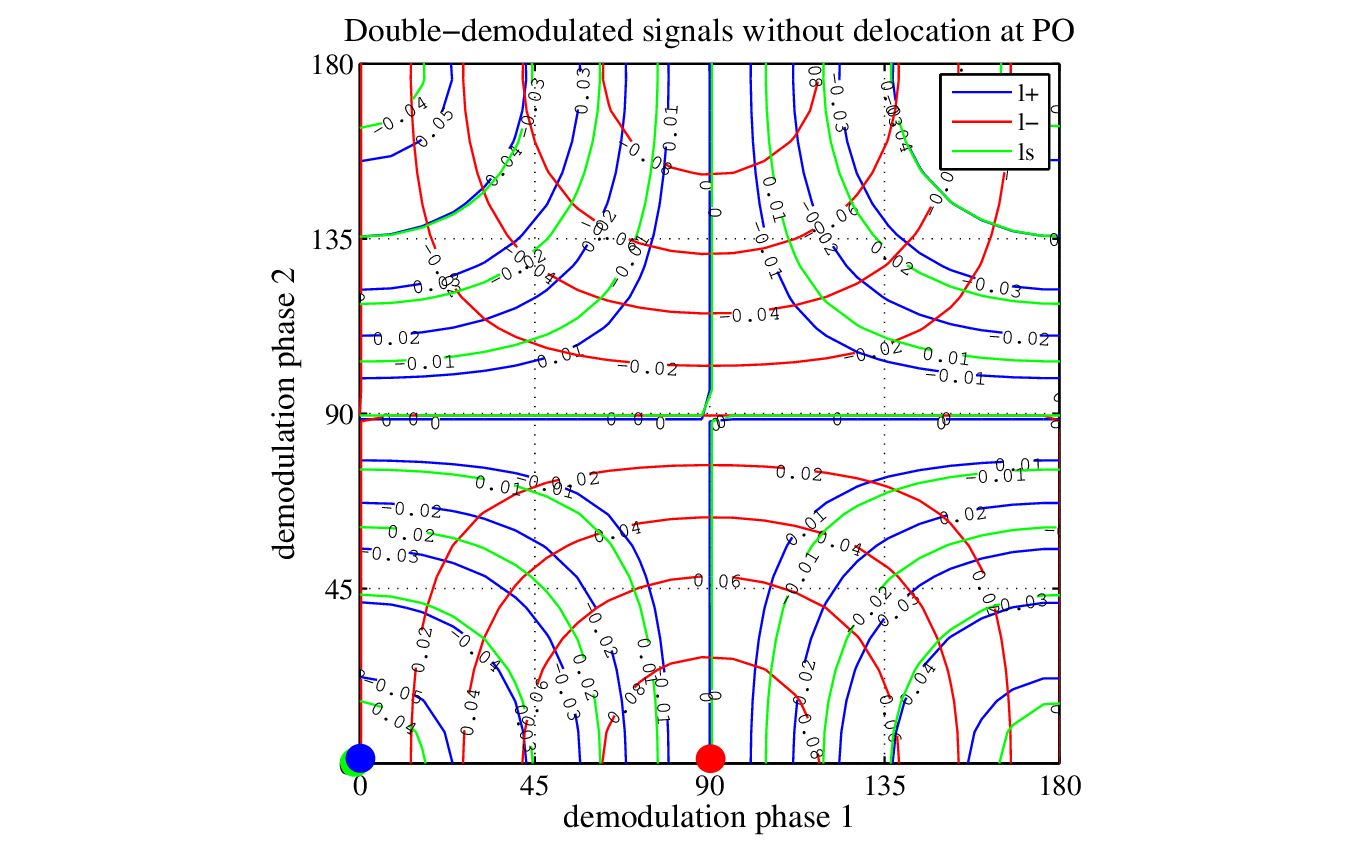}
\vspace{0mm}
\caption{Contour map of $l_+, l_-$ and $l_{\rm s}$ signals
when the delocation is not applied.
The blue, green and red dots represent
the maximum points for the three signals.
At any demodulation phase coordinate,
the maximum points of $l_+$ and $l_{\rm s}$ completely overlap.
They cannot be extracted separately by choosing any demodulation phase.
\\
\\
}
\label{fig:contour1}
%
\includegraphics[width=10cm]{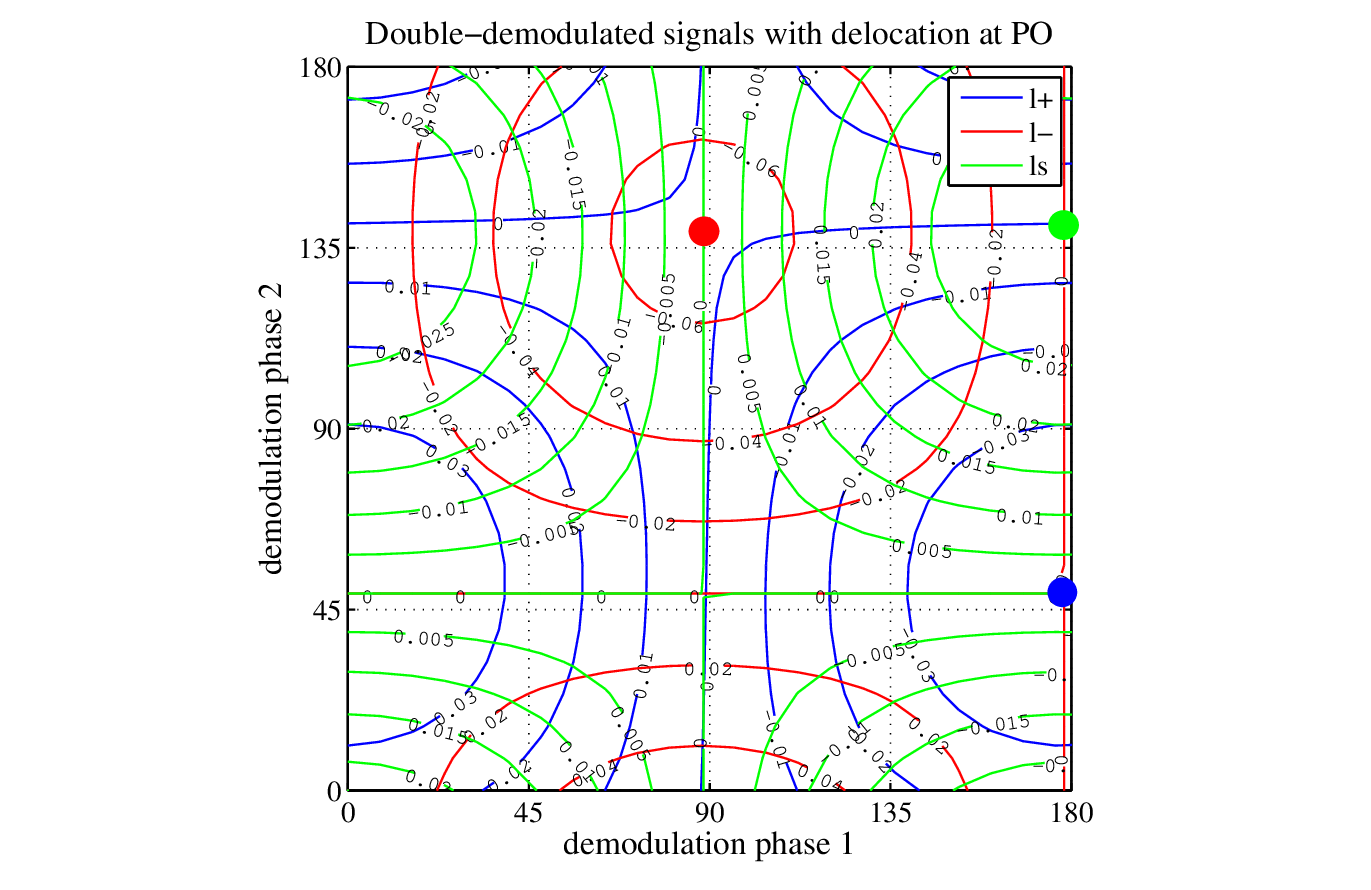}
\vspace{0mm}
\caption{Contour map of $l_+, l_-$ and $l_{\rm s}$ signals
when the delocation is not applied.
The PRM is displaced an appropriate amount from the resonant point.
The delocation length is adjusted so that
undesired signals are almost zero at each maximum point of $l_+, l_-$ and $l_{\rm s}$.
The three signals can be extracted separately at their optimum demodulation phases.}
\label{fig:contour2}
\end{center}
\end{figure}

When the delocation is not applied,
and both the sidebands resonate in the PRC,
$\triangle _{{\rm{p}}1} = \triangle _{{\rm{p}}2} = 0$.
Therefore the signals without the delocation are
%
%
%
\begin{eqnarray}
\label{lpcontour}
\frac{\partial V_{\rm{PO}}}{\partial l_{+}} \propto \cos \delta 1 \cos \delta 2 \\
\frac{\partial V_{\rm{PO}}}{\partial l_{-}} \propto \sin \delta 1 \cos \delta 2 \\
\label{lscontour}
\frac{\partial V_{\rm{PO}}}{\partial l_{{\rm s}}} \propto \cos \delta 1 \cos \delta 2
\end{eqnarray}
Eq.~(\ref{lpcontour}) and Eq.~(\ref{lscontour}) indicate that
$l_{\rm{p}}$ and $l_{\rm s}$ are exactly overlapping
on the $\delta _1-\delta _2$ plane.
The demodulation phases for both $l_{\rm{p}}$ and $l_{\rm s}$
are $(\delta _1,\delta _2)=(0,0)$ to maximize the signals.
Therefore, the two signals cannot be extracted separately
by choosing any demodulation phases.
On the other hand, 
the $l_-$ signal can be extracted at the maximal point,
$(\delta _1,\delta _2)=(\pi /2,0)$ where the other two signals are zero.
The contour plots for these signals are shown in Fig.~\ref{fig:contour1}.

When the delocation is applied,
$\triangle _{{\rm p}1}$ and $\triangle _{{\rm p}2}$ are nonzero. 
These nonzero terms change the signal dependencies on $\delta _{j}$
and they can solve the signal degeneracy.
%
Depicted in Fig.~\ref{fig:contour2},
only one signal can be extracted without mixing with others
by choosing appropriate demodulation phases;
for example,
$l_+$ becomes maximum whereas $l_{\rm s}$ is zero
on a certain pair of demodulation phases
with a suitable amount of delocation.
\begin{table*}
\begin{center}
\begin{tabular}{|c|c|c||c|c|c|c|c|}		\hline 
\multicolumn{8}{|c|}{\textbf{Length sensing matrix}}	\\ \hline\hline
Port&$\delta1$&$\delta2$
&$L_+$   &$L_-$  &$l_+$     &$l_-$    & ~~$l_{\rm s}$		\\ \hline
PO&0& &
1 & $3.83\times10^{-8}$ & $-4.25\times10^{-4}$ & $3.52\times10^{-5}$ & $5.08\times10^{-4}$ 	\\ \hline
DP& 0&
& $-4.65\times10^{-12}$ & 1 & $-1.55\times10^{-18}$ &
 $1.00\times10^{-3}$ & $1.55\times10^{-18}$ \\ \hline
PO&178&49.0
& $-1.23\times10^{-3}$ & $2.55\times10^{-8}$ & 1 & $2.25\times 10^{-7}$ & $-2.41\times 10^{-5}$ \\ \hline
DP& 0 & 136
&$4.65\times10^{-4}$ & $1.24\times10^{-3}$ & $-3.77\times10^{-6}$ & 1 &$-3.89\times10^{-9}$ \\ \hline
PO & 178 & 141
& $1.26\times 10^{-3}$ & $-1.79\times 10^{-8}$ &$8.67\times10^{-7}$ &$5.67\times 10^{-7}$ & 1       	\\ \hline
\end{tabular}
\caption{The numerical results of the length-sensing matrix with
the coupled-reflected method.
Five signals are extracted from each port,
using the demodulation scheme. 
The important feature is that the $l_+, l_-$ and $l_{\rm s}$ signals
are almost separated at the appropriate pairs of demodulation phases, 
($\delta 1$ and $\delta 2$).
We used the preliminary parameter for LCGT;
$r_{\rm p}=0.894$, $r_{\rm s}=0.878$.
According to these reflectivities,
the asymmetry length $\Delta \ell=0.824$m
and the sideband frequencies of $f_1=7$MHz and $f_2=182$MHz
were chosen.
For simplicity, the interferometer was assumed to be lossless.
The delocation length was 0.014m
which corresponds to $\triangle_{\rm p2}=3.06 ^\circ $.
This numerical simulation was done by FINESSE software.
}
\label{matrix}
\end{center}
\end{table*}

%

Table \ref{matrix} shows the matrix of the length sensing signals.
The demodulation phases are optimized
to remove undesired signal components.
E.g., $L_+$ signal is extracted at the BP by the single demodulation on 
a demodulation phase 0 whereas the other four signals are contained.
This simulation was done by FINESSE~\cite{FINESSE}.

This well-diagonalized sensing matrix is
supposed to provide the robust control
because the optical diagonalization enables the servo loop to be simple.
Even without the signal diagonalization,
the interferometer can be controlled by using a signal matrix
which has off-diagonal elements.
But in that case, some additional servo systems
are necessary
to compensate for the off-diagonal elements.
Such systems could add noise to the interferometer.


%


\section{\label{Summary}Summary}
In summary, we have proposed a new length sensing control scheme for an RSE interferometer.
The method has three significant advantages for robust control:
a short asymmetry length of MI is available;
the modulation sideband conditions enable
the $l_-$ to have no cross-talk between $L_+, l_+$ and $l_{\rm s}$ at the DP;
and application of the delocation technique diagonalizes the length-sensing matrix.

\section{\label{acknowledgement}acknowledgements}
This research is supported partly 
by a Grant-in-Aid for Scientific Research on Priority
Areas (415) of the Ministry of Education, Culture, Sports,
Science and Technology, from Japan 
and also National Science Foundation cooperative
agreement No. PHY-0107417 from U.S. funding.

\begin{thebibliography}{999}
\bibitem{footnote1} These ports are traditionally called dark and bright,
although they are not bright or dark for an RSE interferometer.
\bibitem{footnote2}
Although the original method for Advanced LIGO was
to use two sets of PM sidebands instead of AM,
and the carrier field is detuned in the SEC,
it is essentially equivalent to
the transmitted-quasi reflected method.
\bibitem{footnote}When $r_{\rm p} =r_{\rm s}$,
the critical coupling condition for PM
sidebands is satisfied in this method as well.
\end{thebibliography}
\end{document}